\documentclass[cits]{PoS}

\title{The chiral transition as an Anderson transition}

\ShortTitle{The chiral transition as an Anderson transition}

\author{Matteo Giordano\llap{}\thanks{Supported by the Hungarian Academy of Sciences under ``Lend\"ulet'' grant No. LP2011-011.}\\
Institute for Nuclear Research of the Hungarian Academy of Sciences\\Bem t\'er 18/c, H-4026 Debrecen, Hungary\\
E-mail: \email{giordano@atomki.mta.hu}
}

\author{S\'andor D. Katz\\MTA-ELTE Lend\"ulet Lattice Gauge Theory Research Group, E\"otv\"os University\\
P\'azm\'any P.~s\'et\'any 1/A, H-1117 Budapest, Hungary\\
E-mail: \email{katz@bodri.elte.hu}}

\author{Tam\'as G. Kov\'acs$^*$\\
Institute for Nuclear Research of the Hungarian Academy of Sciences\\Bem t\'er 18/c, H-4026 Debrecen, Hungary\\
E-mail: \email{kgt@atomki.mta.hu}}

\author{\speaker{Ferenc Pittler}\llap{}\thanks{Supported by OTKA under the grant OTKA-NF-104034.}\\MTA-ELTE Lend\"ulet Lattice Gauge Theory Research Group, E\"otv\"os University\\
P\'azm\'any P.~s\'et\'any 1/A, H-1117 Budapest, Hungary\\
E-mail: \email{pittler@bodri.elte.hu}}

\abstract{At low temperature the low-lying QCD Dirac 
spectrum obeys random matrix statistics. Recently we 
found that above $T_{c}$ the lowest part of the spectrum 
consists of localized modes that obey Poisson statistics. 
An interesting implication of this is that as the system 
crosses $T_{c}$ from above, the spectral statistics at 
$\lambda=0$ changes from Poisson to random matrix. 
Here we study this transition and its possible implications 
for the finite temperature transition of QCD-like theories.}

\FullConference{The 32nd International Symposium on Lattice Field Theory,\\
		23-28 June, 2014\\
		Columbia University New York, NY}

\begin{document}

\section{Introduction}
It is well known that at low temperature the low-lying eigenmodes 
of the QCD Dirac operator are delocalized, and their statistics 
can be described by random matrix theory \cite{Shuryak:1992pi,BerbenniBitsch:1997tx}. 
This fact is utilized also in practice: low energy constants 
of chiral perturbation theory can be determined by comparing 
lattice data and random matrix predictions. Recently it was found 
that the high-temperature Dirac spectrum is more ``complex'' than 
the low temperature one \cite{GarciaGarcia:2005vj,Kovacs:2009zj,Kovacs:2010wx,Kovacs:2012zq}.
This means that the eigenmodes of the Dirac operator are either 
localized or delocalized depending on where they are located in 
the spectrum. The low end of the spectrum consists of localized, 
statistically uncorrelated eigenmodes, while modes in the bulk 
still remain delocalized and statistically correlated. The QCD 
Dirac operator is not the only physical system having such 
a ``complex'' spectrum. In solid state physics, the Hamiltonian 
of the Anderson model provides an example of this kind of spectrum. 
The Anderson model is the simplest model for Anderson localization
\cite{Anderson:1958vr}. The one-electron Hamiltonian of the Anderson 
model describes non-interacting electrons and has the form
\begin{equation}
\label{andrtight}
H= \sum_{i} \varepsilon_{i} \vert i \rangle \langle i \vert +  
\sum_{\langle i,j\rangle}\vert i \rangle \langle j\vert,
\end{equation}
where ${i}$ represents an atomic orbital on a lattice site. The 
first term in eq.~(\ref{andrtight}) represents an on-site 
interaction with randomly distributed site energies $\epsilon_{i}$.  
The second term is responsible for nearest-neighbor hopping with
constant coupling strength. In the case when the disorder is zero,
i.e., the variance $\sigma\left(\varepsilon_{i}\right)$ of the
random on-site energies is zero, the eigenfunctions of $H$ are 
the well-known, delocalized Bloch wavefunctions. 
If the disorder is turned on then localized eigenmodes will appear 
around the spectrum edges, while states near the band center remain 
delocalized, and there will be a boundary between the localized and 
delocalized modes in both the positive and negative part of the 
spectrum called mobility edge. If the disorder increases the mobility 
edges move further towards the band center, and eventually at 
a critical disorder strength all eigenmodes of $H$ will be 
localized. This transition is known as the Anderson transition.

The Anderson transition in solid-state physics is a real second order 
phase transition \cite{lee}. Some of us found that the finite 
volume crossover in the high-temperature Dirac spectrum from localized 
eigenmodes to delocalized eigenmodes also becomes a genuine second-order 
phase transition in the thermodynamic limit \cite{Giordano:2013taa}. In 
the same work we also determined the critical exponent 
$\nu_{QCD}=1.43\left(6\right)$ which is consistent with the one found in 
the three dimensional unitary Anderson model $\nu_{And}=1.43\left(4\right)$
\cite{andcritexp}. A possible explanation for this consistency, and a more 
detailed description of the similarities of the spectra of the Anderson 
model and of the QCD Dirac operator can be found in \cite{giordano2014}. 
In the Anderson model the control parameter which drives the transition 
is the disorder strength $\sigma\left(\epsilon_{i}\right)$. On the other 
hand, the transition can be also examined at fixed disorder strength 
by scanning the spectrum. Physically, this approach arises more naturally, 
since one can control the position of the Fermi energy by changing the 
chemical potential and examine the transition as the Fermi energy crosses 
the mobility edge. Unfortunately, in QCD (as far as we know) there is no 
physical control parameter which enables us to change the position in 
the Dirac spectrum. In fact, the mobility edge is only a point in the 
spectrum, while all local thermodynamical quantities (for example the 
quark condensate $\langle \bar{\psi} \psi \rangle$) are obtained by 
integrating over the full spectrum. Nevertheless, the presence of 
localized low-lying modes might affect these observables.
 
In lattice QCD the control parameter is the temperature, which plays a 
similar role to the disorder strength $\sigma\left(\epsilon_{i}\right)$ 
in the Anderson model. In this paper we analyze the localization properties 
of low-lying modes as the temperature is varied, both in QCD and in a QCD-related 
toy model displaying a true first order phase transition. We also study the 
relation between the appearance of localized modes and the presence of a 
dramatic change in the thermodynamic properties of the system.

\section{Phase diagram of QCD}
\begin{figure}
\begin{center}
\includegraphics[scale=0.5]{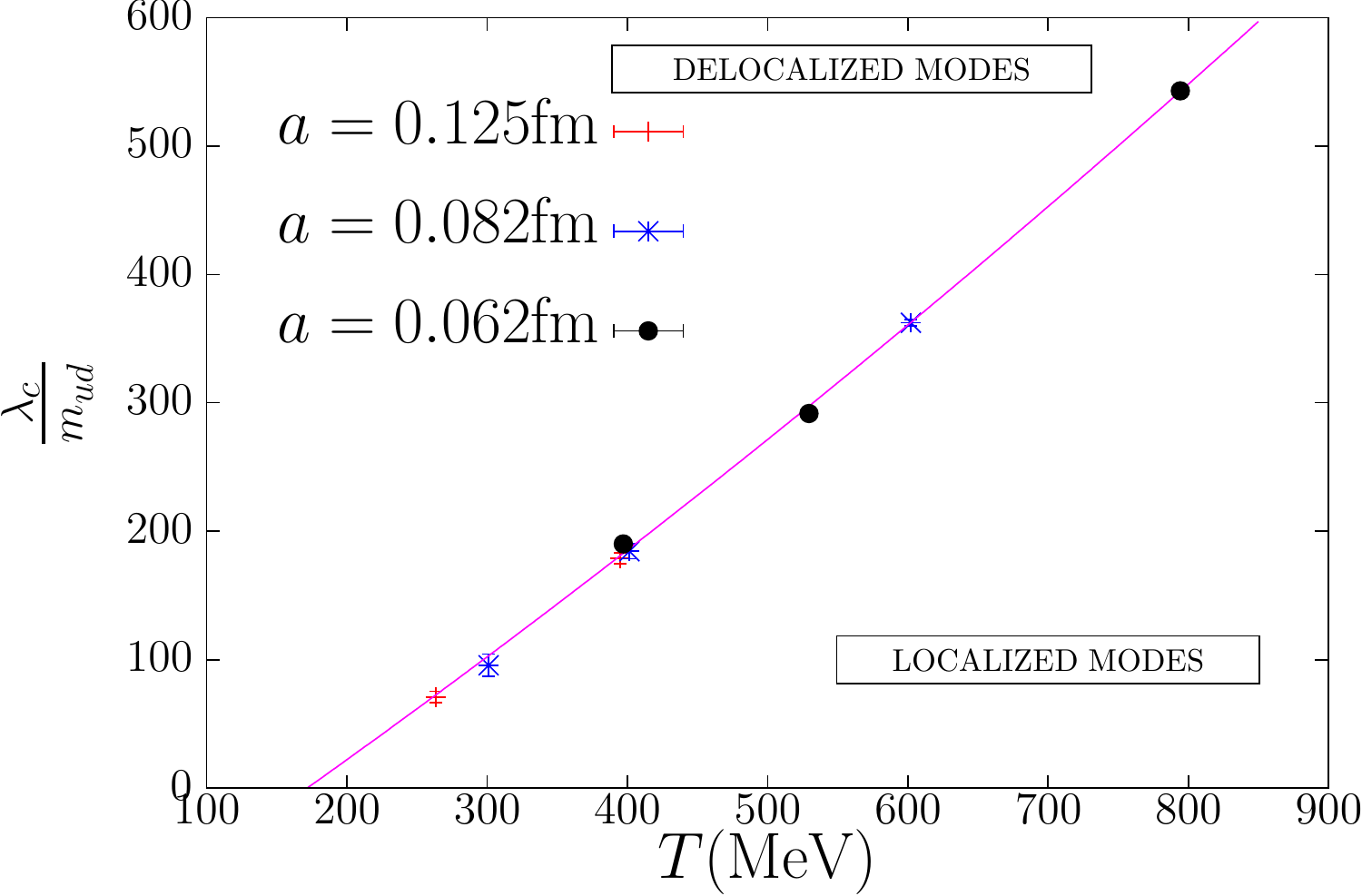}
\caption{Phase diagram of $QCD$ in the eigenvalue-temperature plane.}
\label{qcdphasediagram}
\end{center}
\end{figure}

To analyze the temperature dependence of the mobility edge, $\lambda_c$, 
the appropriate quantity to consider is its ratio to the bare quark mass. 
Since $\lambda_c$ acts effectively as a bare quark mass, it is expected to 
renormalize similarly, so that $\lambda_c/m_{ud}$ is renormalization group 
invariant.
A more formal argument can be found in \cite{Kovacs:2012zq}. 
In Fig.~\ref{qcdphasediagram} we show the phase diagram of QCD in 
the $\left(\lambda,T\right)$ plane. To a good precision, at high 
temperature, well above the pseudo-critical QCD transition temperature 
$T_{c}$, the mobility edge depends linearly on the temperature. There 
is a critical line in the phase diagram along which there is a second 
order phase transition in the spectrum. 
If we extrapolate this critical line to vanishing mobility edge, 
we get a temperature consistent with the pseudo-critical QCD crossover 
temperature. However, the relationship  between the Anderson transition 
and the QCD chiral crossover is not straightforward, because in the latter 
case there is no unique transition temperature. In order to investigate 
the relationship between the chiral transition and the Anderson transition 
in a clear-cut setting we will consider a simple model which shows a real 
first order phase transition in the thermodynamic limit. In fact our work 
has also practical motivations and might suggest new order parameters for 
locating the chiral transition. Usually, the chiral transition is analyzed
by comparing the value of the quark condensate at different temperatures. 
A priori, calculating the bare quark condensate at just one temperature will 
not tell us to which phase the ensemble of configurations belongs.
In this work our aim is also to look for quantities that can be easily 
computed (without any renormalization or zero temperature simulation) 
and can be used to identify the phase to which the ensemble of 
configurations belongs.
\section{A toy model: three flavors of unimproved staggered quarks on $N_{t}=4$ lattices}
In order to investigate the relationship between 
the chiral transition and the Anderson transition 
in the spectrum we simulate $N_{f}=3$ unimproved 
staggered quarks with the Wilson plaquette action. At $N_{t}=4$ 
this toy model has a real first order phase transition 
in the thermodynamic limit for sufficiently small 
quark masses \cite{deForcrand:2008vr}. In our case we 
use $m_{bare}=0.01$. To check the finite volume effects 
we use three different lattice volumes: $24^{3},32^{3},48^{3}$.
Since $N_{t}$ is fixed we use the $\beta$ gauge coupling to 
change the temperature.  
Although, from the point of view of QCD, this first order 
phase transition is a coarse lattice artifact, our toy model 
is nevertheless a legitimate statistical physics model, 
which allows to compare, in a clear-cut setting, the temperature 
at which a genuine phase transition occurs, and the temperature 
at which localized modes appear in the Dirac spectrum.


\section{Numerical results}
Let us first check how the spectral density changes 
when the first order transition happens, and then to examine 
how the spectral statistics change along the spectrum at 
temperatures which definitely belong to the ``chirally symmetric'' phase. 
Establishing the transition in the spectrum in the 
``chirally symmetric'' phase we define a mobility edge and 
map out the phase diagram of our toy model in the 
$\left(\lambda,\beta\right)$ plane. 

The usual way to examine a chiral transition is to study 
the temperature dependence of the quark condensate. The latter can be
expressed in terms of the spectral density of the lattice Dirac operator 
with the help of the Banks-Casher formula \cite{Banks:1979yr} 
as follows,
\begin{equation}
\langle \bar{\psi} \psi\left(m\right) \rangle = 
\int_{0}^{\infty} \mathrm{d} \lambda \frac{2m}{\lambda^{2}+m^{2}}\rho\left(\lambda\right).
\end{equation}
A sizeable contribution to the integral comes from the low modes 
of the Dirac operator, so a large change in the spectral density 
around zero will reflect in a large change of the condensate. 
\begin{figure}
\begin{center}
\includegraphics[scale=0.5]{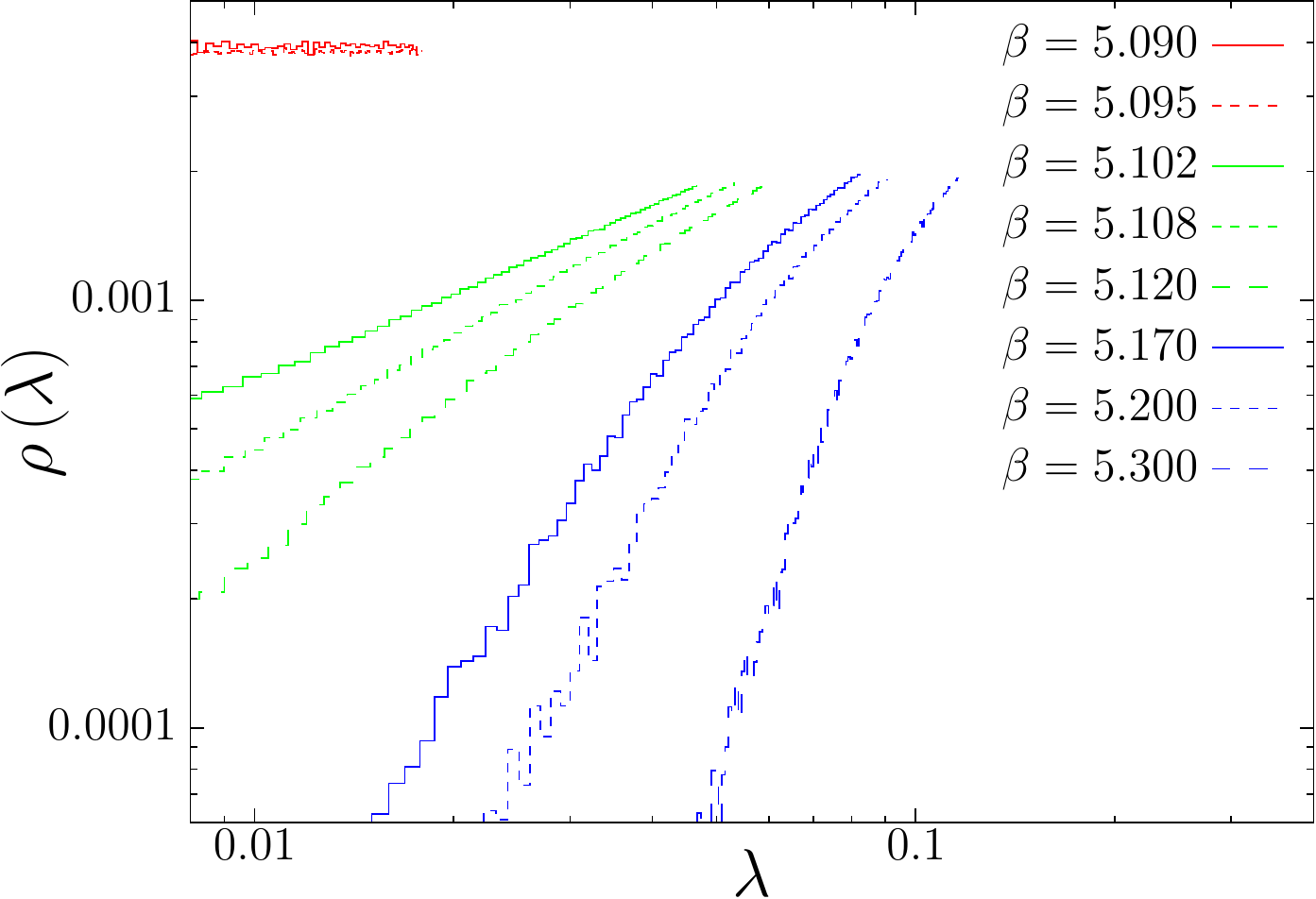}
\caption{Spectral density (normalized by the lattice volume) of our toy model 
for different values of $\beta$ on a log-log scale.}
\label{sdensity}
\end{center}
\end{figure}
In Fig.~\ref{sdensity} we show how the spectral density near zero 
changes as $\beta$ is varied and the system is driven through the chiral 
phase transition. Below $\beta=5.099$ the spectral density near the origin 
is essentially constant in $\lambda$, which leads to a large contribution 
to the chiral condensate, and changes only slightly with $\beta$. Above 
$\beta=5.099$ the spectral density vanishes continuously as a power law 
approaching the origin, so the contribution of the low modes to the condensate 
drops considerably. As a matter of fact, the chiral condensate has a jump 
around $\beta=5.099$ (see Fig.~\ref{phasetoy}), so that the two different 
behaviors of $\rho$ near the origin correspond to the two phases of the system.

The presence of localized modes in the spectrum can be conveniently 
detected by analyzing the spectral statistics. Localized modes usually 
appear in a region of low spectral density, and so they are unlikely 
to be mixed by the gauge field fluctuations, as they have large eigenvalue 
differences and small eigenmode overlaps. They are therefore expected 
to be statistically independent, and the corresponding eigenvalues should 
follow Poisson statistics. On the other hand, when the spectral density 
is large the fluctuations can easily mix the modes, leading to delocalised 
and correlated eigenmodes, with eigenvalues distributed according to random matrix
theory. 

In order to determine what kind of statistics the eigenvalues 
obey, it is easier to work with the so-called unfolded spectrum,
obtained by properly rescaling the eigenvalues to have unit spectral
density throughout the spectrum. In practice, this is done by
replacing each eigenvalue with its rank in the full sequence of
eigenvalues.
A commonly used spectral statistics is the so-called unfolded level spacing 
distribution ($ULSD$) $p\left(s\right)$. In the case of uncorrelated 
eigenvalues the $ULSD$ is an exponential distribution: $p_{Poisson}\left(s\right) 
= \exp\left(-s\right)$, and in the case of correlated eigenvalues the $ULSD$ 
can be well approximated by the so-called Wigner surmise of the corresponding 
random matrix ensemble $p_{RMT}\left(s\right)=\frac{32}{\pi^{2}}s^{2}
\exp\left(-\frac{4}{\pi}s^{2}\right)$. First we check whether the low modes of 
the high-temperature Dirac spectrum are also statistically independent in our toy model. 
To show this we choose some parameter of the $ULSD$, compute it locally 
in the spectrum and monitor how it changes between the two limiting cases. 
We choose the integral of the $ULSD$ up to the point where the exponential and 
the Wigner surmise first cross (approximately):
\begin{equation}
I_{0.5}=\int_{0}^{0.5}\mathrm{d}s~p\left(s\right).
\end{equation}
\begin{figure}
\begin{center}
\includegraphics[scale=0.5]{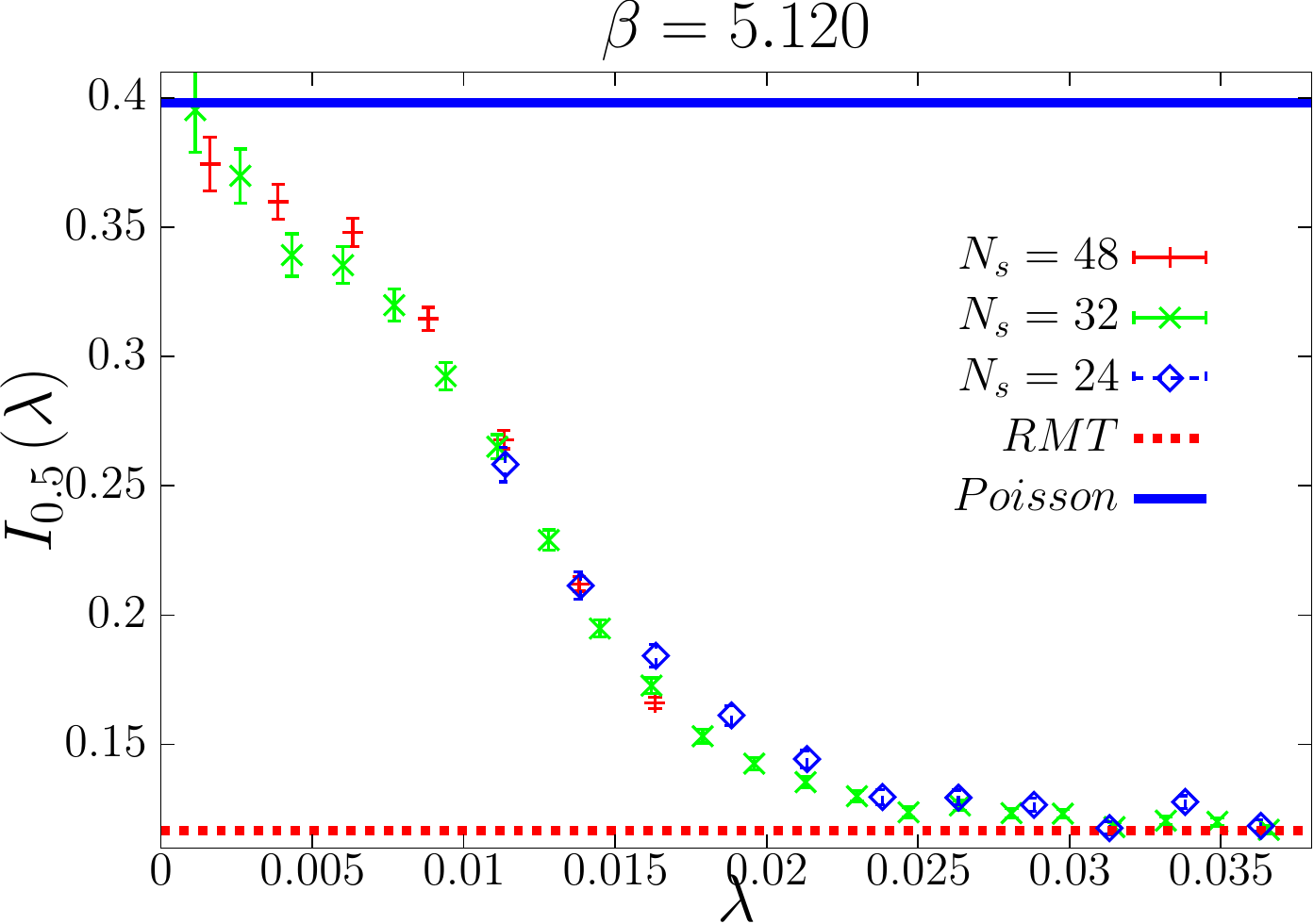}
\includegraphics[scale=0.5]{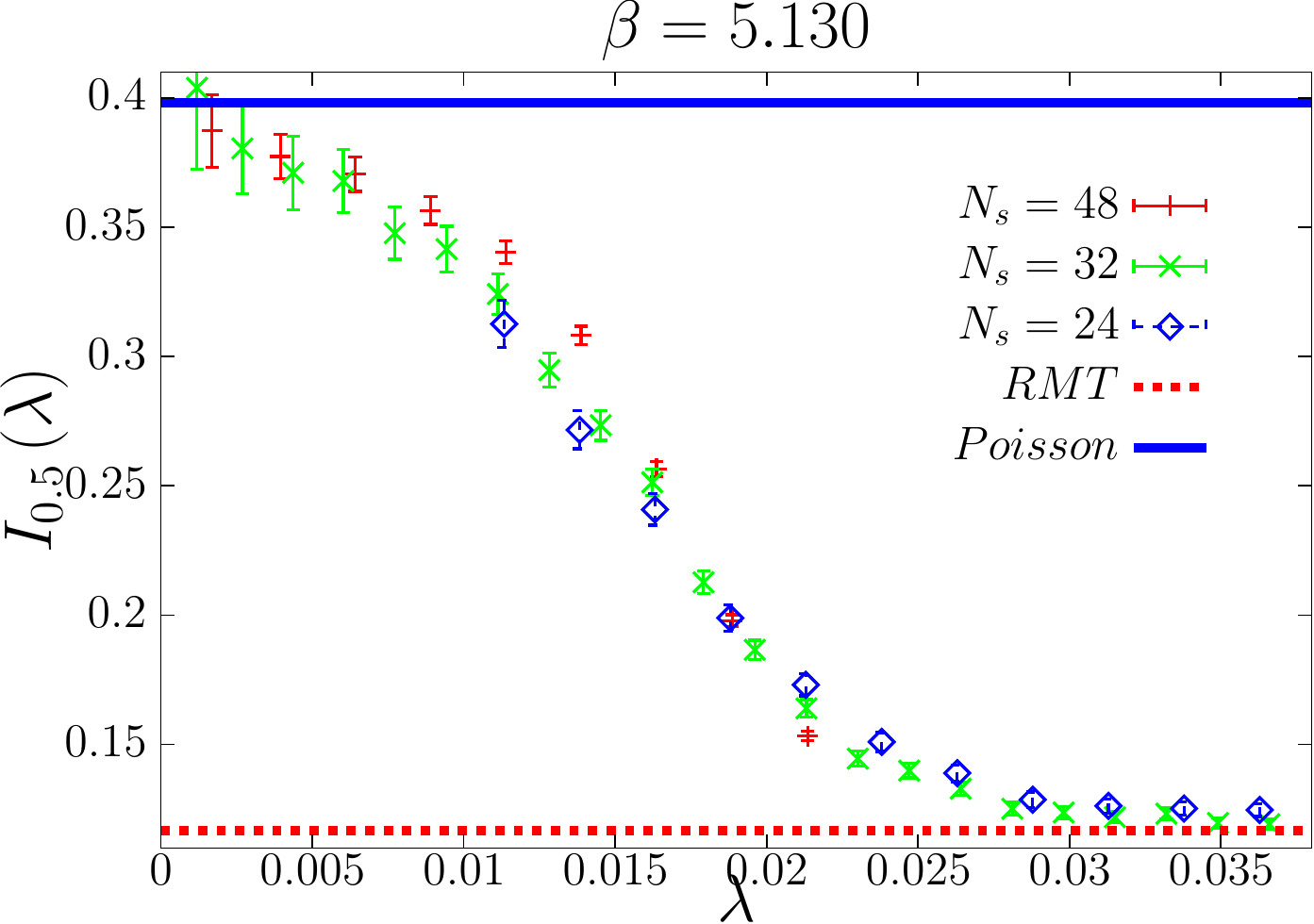}
\caption{Poisson-RMT transition in high-temperature Dirac spectra 
for two different temperatures.\label{i05beta}}
\end{center}
\end{figure}
In Fig.~\ref{i05beta} we show how this statistic changes locally 
in the spectrum for two different temperatures. For both temperatures 
we see qualitatively the same behavior: moving 
from the edge towards the bulk of the spectrum the spectral statistics 
changes from Poisson to random matrix. For each temperature we 
show the continuous transition in the spectrum for three different 
lattice volumes. It is clear that the transition becomes sharper on 
larger volumes, suggesting a singular behavior in the thermodynamic 
limit. For the QCD Dirac spectrum we showed that this transition is 
a second order phase transition. Most likely this is the case for our 
toy model too, since around the critical point in the spectrum 
(the mobility edge), where $I_{0.5}$ is volume independent,
this quantity takes the same value as in QCD, $I_{0.5}\left(\lambda_{c}\right)=0.187$
\cite{Nishigaki:2013uya}, both for $\beta=5.120$ and $5.130$. We also 
see from Fig.~\ref{i05beta} that the mobility edge decreases as the 
temperature decreases. In the following we examine how this mobility 
edge goes to zero as a function of the temperature and how this correlates 
with the location of the first order chiral phase transition.

\section{Phase diagram of the toy model}

\begin{figure}
\begin{center}
\includegraphics[scale=0.6]{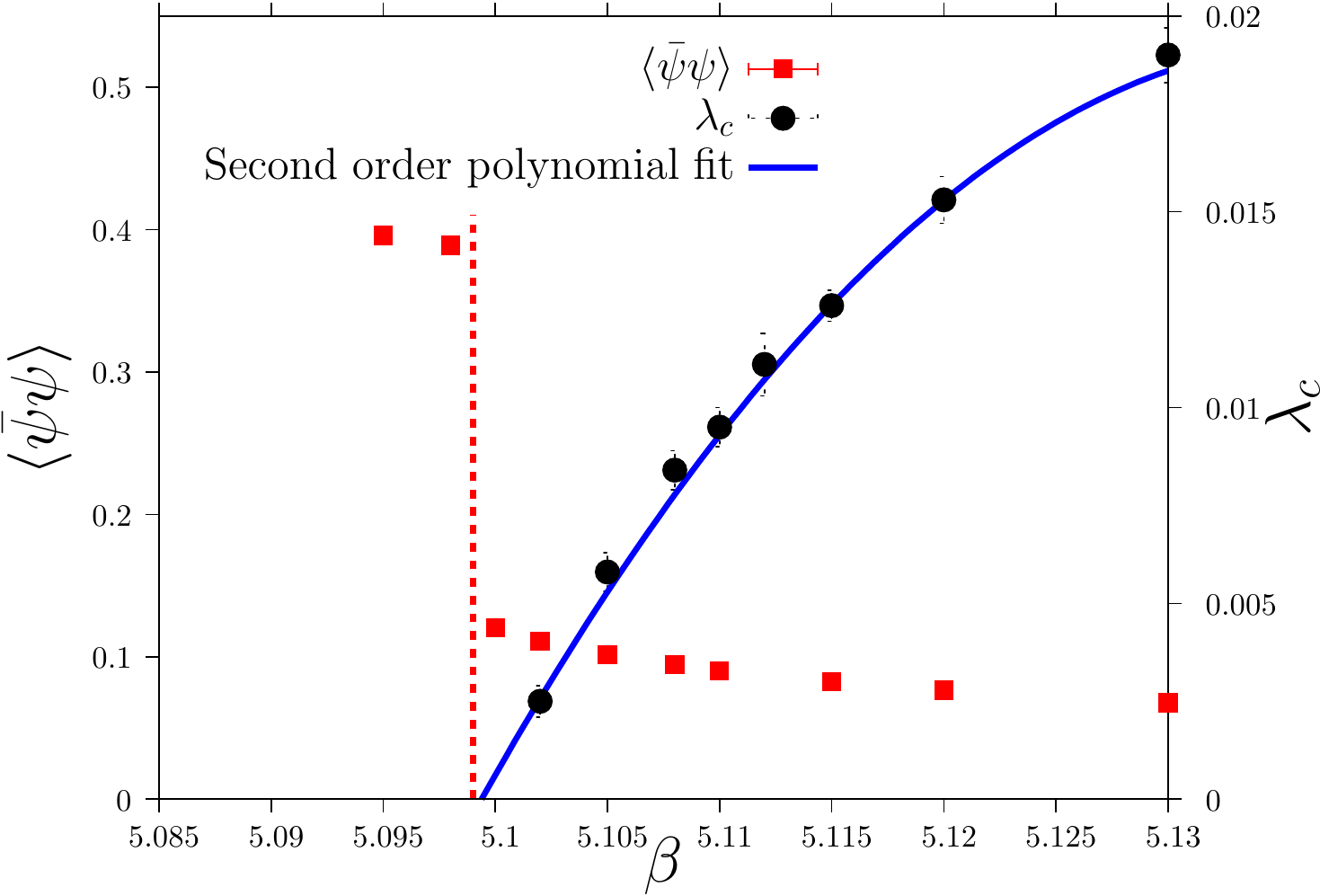}
\caption{The phase diagram of our toy model in the $\left(\lambda,\beta\right)$ 
plane. We show the mobility edge with filled circles and the value of 
$\langle\bar{\psi}\psi\rangle$ with filled squares. The smooth line represents the 
critical line in the $\left(\lambda,\beta\right)$ plane. The vertical dotted line corresponds 
to $\beta_{c}=5.099$ where the (extrapolated) mobility edge vanishes.}
\label{phasetoy}
\end{center}
\end{figure}

For any value of $\beta$, the mobility edge can be identified as 
the point in the spectrum where local spectral statistics are
volume-independent in the large-volume limit.  
With decreasing $\beta$ the mobility 
edge goes towards the spectrum edge and we need larger and larger 
volumes to capture the volume dependence of the transition. 
At large lattice volume, however, we can determine the mobility
edge using simulations at a single volume only.
We identify the mobility edge with the location in the spectrum where 
the spectral statistics takes its critical value i.e. $I_{0.5}\left(\lambda\right)=0.187$. 
In Fig.~\ref{phasetoy} we show how the mobility edge changes as a function of
the temperature. 
In the same figure we also show a second-order polynomial fit to
the data points. The fit can be used to read off the critical 
$\beta_{c}$ where the mobility edge becomes 
zero. In the same figure we also show how $\langle \bar{\psi}\psi\rangle$ 
changes as a function of $\beta$. We calculated 
$\langle \bar{\psi}\psi\rangle$ with the stochastic method. This quantity, 
being the first derivative of the partition function, exhibits a 
discontinuity at a real first order phase transition. Our most 
important result is that the mobility edge becomes zero around 
the point $\beta_{c}=5.099$, where the value of the $\langle\bar{\psi}\psi\rangle$ 
jumps. This means that the localized modes appear at the 
temperature  where the chiral transition occurs. 

\section{Conclusions and outlook}
The question whether the effect of the chiral crossover in QCD can 
be seen also in terms of localization of the eigenmodes of the Dirac 
operator is a rather old one \cite{GarciaGarcia:2005vj}. At that time 
high statistics simulations with physical quark masses were unfeasible. 
In QCD with physical quark masses we found that the Anderson transition 
in the Dirac spectrum starts to appear around 163~MeV \cite{tamas}, which 
is consistent with the pseudo-critical QCD transition temperature determined 
from other quantities \cite{borsanyi}. In this paper we deal with a model 
in which there is a real first order phase transition in the thermodynamic 
limit. Our main result is that in this model the chiral transition and the 
Anderson transition occur at the same temperature.  The next step would be 
to understand if and how the Anderson transition can possibly explain the 
chiral transition. To this extent, it would be interesting to extrapolate 
the spectral statistics to $\lambda=0$ at a given value of $\beta$ at a fixed 
lattice volume $V$ and then do a finite-size scaling analysis. 
However, to get a reliable extrapolation to $\lambda=0$ around $\beta_{c}$ we need 
even larger volumes. Another interesting possibility is to study the statistical
behavior of the first and the second eigenvalue of the lattice Dirac operator. 
For that aspect the overlap Dirac operator would be much better, since it 
does not suffer from taste-symmetry breaking of staggered fermions. 

\end{document}